\documentclass[prd,aps,floats,
twocolumn,tightenlines,notitlepage]{revtex4}

\usepackage{epsfig}

\def\be{\begin{equation}}
\def\ee{\end{equation}}
\def\beq{\begin{eqnarray}}
\def\eeq{\end{eqnarray}}
\def\a{\alpha}
\def\b{\beta}

\def\vp{\varphi}

\def\N{{\cal N}}
\def\O{{\cal O}}

\def\V{{\cal V}}

\def\({\left (}
\def\){\right )}
\def\[{\left [}
\def\[{\right ]}

\begin{document}
\title{Designer Gravity and Field Theory Effective Potentials}

\author{Thomas Hertog and Gary T. Horowitz}

\affiliation{Department of Physics, UCSB, Santa Barbara, CA 93106}

\bibliographystyle{unsrt}

\begin{abstract}
Motivated by the AdS/CFT correspondence, we show that there is a
remarkable agreement between static supergravity solutions and
extrema of a field theory potential. For essentially  any function
$\V(\a)$ there are boundary conditions in anti de Sitter space so
that gravitational solitons exist precisely at the extrema of $\V$
and have masses given by the value of $\V$ at these extrema. Based
on this, we propose new positive energy conjectures. 
On the field theory side, each function $\V$ can be interpreted as the 
effective potential for a certain operator in the dual field theory.

\end{abstract}
%\end{titlepage}
\maketitle

{\bf Introduction.}
In theories of gravity coupled to matter, the theory is usually fully
determined by the action. The boundary conditions at
infinity are often not independent, but uniquely determined by basic
requirements such as finite total energy. This is
not the case, however, for certain theories of gravity in asymptotically
anti de Sitter (AdS) spacetimes. For the same action, there can be 
many possible boundary conditions, and changing the boundary
conditions changes the properties of the theory. In particular,
we will see that one can ``pre-order" the number and masses of 
solitons in some theories: There are boundary conditions which yield 
any desired result. For this reason, we call this phenomenon 
``designer gravity".

Among the theories of gravity for which this is possible are
certain supergravity theories. In fact, although this result is
independent of string theory, it was discovered
while investigating the AdS/CFT correspondence \cite{Maldacena98}.
Furthermore, in cases where there is a field theory dual, the
gravitational solitons can be used to compute certain effective potentials 
in the field theory.

We will consider theories of gravity coupled to a scalar field
with potential $V(\phi)$. We require that $V$ has a negative
maximum, so that AdS is a solution and small scalar fluctuations
are tachyonic, $m^2<0$.   It has long been known that tachyonic
scalars in $d+1$ dimensional AdS spacetime  are stable provided
their mass is above the Breitenlohner-Freedman (BF) bound
\cite{Breitenlohner82} $m^2_{BF} = -d^2/4$ (in units of the AdS
radius). It has been shown much more recently that if
\be\label{bound} m_{BF}^2 \le m^2 <m_{BF}^2 +1 \ee then more
general boundary conditions are possible which still admit a
conserved finite total energy and preserve all the AdS symmetries
\cite{Hertog04,Henneaux04}.

For definiteness, we will  focus on the case of $\N=8$ gauged
supergravity in four dimensions, and comment on generalizations at
the end.
 This theory can be consistently truncated to
 include just gravity and a single scalar field with potential
 \cite{Duff99}
 \be\label{pot}
 V(\phi) = -2-\cosh(\sqrt{2}\phi)
 \ee
 so, setting $8\pi G=1$, our action is
 \be \label{4-action} S=\int d^4x\sqrt{-g}\left[\frac{1}{2}R
-\frac{1}{2}(\nabla\phi)^2 +2+\cosh(\sqrt{2}\phi) \right] \ee
 The potential (\ref{pot}) has a maximum at $\phi=0$ corresponding to 
 an $AdS_4$ solution with unit
 radius.  It is unbounded from below, but small fluctuations have
 $m^2 =-2$ which is above the BF bound, and satisfies (\ref{bound}).

In all asymptotically AdS solutions, the scalar $\phi$  decays
at large radius as
 \be\label{hair4d}
  \phi(r)=\frac{\alpha}{r}+\frac{\beta}{r^2}
  \ee
where $r$ is an asymptotic area coordinate, and $\a$ and $\b$ can
depend on the other coordinates. The standard boundary conditions
correspond to either $\a=0$ or $\b=0$ \cite{Breitenlohner82,Klebanov:1999tb}.
 It was shown in \cite{Hertog04} that $\b = k\a^2$
(with $k$ an arbitrary constant) 
was another possible boundary condition that preserves all the
asymptotic AdS symmetries. We now consider even
more general boundary
conditions $\b = \b(\a)$. Although these will generically break
some of the asymptotic AdS symmetries, they are invariant under
global time translations. Hence there is still a conserved total
energy, as we now show.

As discussed in \cite{Hertog04}, the usual definition of energy in
AdS diverges whenever $\a\ne 0$. This is because the 
backreaction of the scalar field causes certain 
metric components to fall off slower than usual. The
complete set of boundary conditions can be found in
\cite{Hertog04} but the main change is in $g_{rr}$:
 \be \label{4-grr}
g_{rr}=\frac{1}{r^2}-\frac{(1+\alpha^2/2)}{r^4}+
O(1/r^5)
\ee

The expression for the conserved
mass depends on the asymptotic behavior 
of the fields and is defined as follows.
Let $\xi^\mu$ be a timelike vector which asymptotically approaches
a (global)  time translation in AdS. The Hamiltonian takes the
form
\be
 H = \int_\Sigma \xi^\mu  C_\mu + {\rm surface \ terms}
\ee
 where $\Sigma$ is a spacelike surface, $C_\mu$ are the usual
constraints, and the surface terms should be chosen so that the
variation of the Hamiltonian is well defined. The variation of the
usual gravitational surface term is given by
\beq\label{gravch}
\delta Q_{G}[\xi]&=&\frac{1}{2}\oint dS_i
\bar G^{ijkl}(\xi^\perp \bar{D}_j \delta h_{kl}-\delta h_{kl}\bar{D}_j\xi^\perp)
\eeq
where $G^{ijkl}={1 \over 2} g^{1/2} (g^{ik}g^{jl}+g^{il}g^{jk}-2g^{ij}g^{kl})$,
$h_{ij}=g_{ij}-\bar{g}_{ij}$ is the deviation from the spatial metric 
$\bar{g}_{ij}$ of pure AdS,  $\bar{D}_i$ denotes covariant differentiation 
with respect to $\bar{g}_{ij}$ and $\xi^\perp = \xi \cdot n$ with $n$ the
unit normal to $\Sigma$.  Since our scalar
field is falling off more slowly than usual if $\a \ne 0$, 
there is an additional scalar contribution to the surface terms. 
Its variation is simply
\be\label{deltaQ}
\delta Q_\phi[\xi] =-\oint \xi^\perp \delta \phi D_i \phi dS^i \ee
 Using the asymptotic
behavior (\ref{hair4d})  this becomes
\be
\delta Q_\phi[\xi] = r\oint(\a\delta \a)d\Omega 
+ \oint [\delta (\a\b) + \b\delta \a]d\Omega
\ee
Since there is a
term proportional to the radius of the sphere, this scalar surface
term diverges. However, this divergence is exactly canceled by the
divergence of the usual gravitational surface term (\ref{gravch}).
The total charge can therefore be integrated, yielding
\be\label{scalarst}
Q [\xi] = Q_{G}[\xi]+ r\oint {\a^2\over 2}d\Omega  + 
\oint [\a\b + W(\a)]d\Omega \ee
where we have defined
\be W(\a) = \int_0^\a \b(\tilde \a)d\tilde \a \ee
In addition to canceling the divergence in (\ref{scalarst}), the
gravitational surface term contributes a finite amount $M_0$. For
the spherically symmetric solutions we consider below, this is
just the coefficient of the $1/r^5$ term in $g_{rr}$. Since $\a$
and $\b$ are now independent of angles, the total mass becomes
\be\label{mass}
M= 4\pi(M_0 + \a\b + W)
\ee
(For $\b=k\a^2$, this agrees with the expression for the mass given in
\cite{Hertog04}.)

{\bf Gravitational Solitons.}
We want to study solitons in this theory. These are nonsingular,
static, spherically symmetric solutions. Writing the metric as \be
ds^2=-h(r)e^{-2\chi(r)}dt^2+h^{-1}(r)dr^2+r^2d\Omega \ee the
field equations read \be\label{hairy14d}
h\phi_{,rr}+\left(\frac{2h}{r}+\frac{r}{2}\phi_{,r}^2h+h_{,r}
\right)\phi_{,r}   =  V_{,\phi} \ee \be\label{hairy24d}
1-h-rh_{,r}-\frac{r^2}{2}\phi_{,r}^2h =  r^2V(\phi) \ee \be
\chi_{,r} = -{r\phi_{,r}^2\over 2} \ee
 Regularity at the origin requires $h=1$ and $ h' = \phi'= \chi'=0$. Rescaling $t$ shifts
 $\chi$ by a constant, so its value at the origin is arbitrary. Thus solutions can be labeled by
 the value of  $\phi$ at the origin. For each $\phi(0)$,
 one can integrate these ODE's and get a soliton.
Asymptotically, $\phi$ behaves like (\ref{hair4d}), so we get a
point in the $(\a,\b)$ plane. Repeating for all  $\phi(0)$ yields
a curve $\b_s(\a)$ where the subscript indicates this is
associated with solitons. This curve is plotted in Fig.
1. (Since the potential $V(\phi)$ is even, it suffices to
consider positive $\phi(0)$ which corresponds to positive $\a$.)
Note that solitons exist for arbitrarily small $\a$. 
When $\a\ll 1$, $\phi(r)$ is small everywhere, and one might have
thought a linearized approximation should be valid implying no solitons
could exist. This is incorrect since for any $\a\ne 0$, the backreaction 
is always large asymptotically as shown in (\ref{4-grr}).
Given a choice of boundary condition $\b(\a)$, the allowed
solitons are simply given by points where the soliton curve
intersects the boundary condition curve: $\b_s(\a) = \b(\a)$.

%%%%%%%%%%%%%%%%%%%%%%%%%%%%%%%
\begin{figure}[htb]
\begin{picture}(0,0)
\put(0,60){$\beta$} 
\put(210,140){$\alpha$}
\end{picture}
\psfig{file=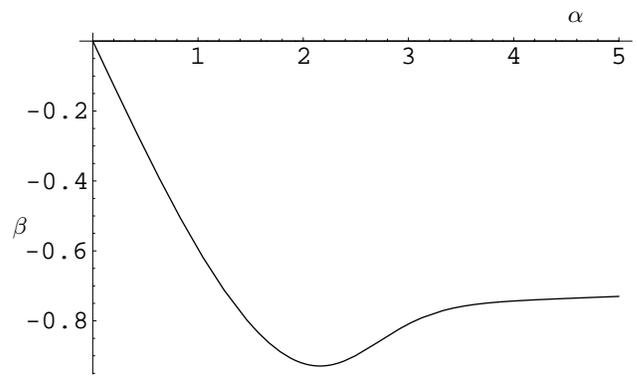,width=3.2in} 
\caption{The function $\beta_s$  obtained from the solitons.} 
\label{1}
\end{figure}
%%%%%%%%%%%%%%%%%%%%%%%%%%%%%%%%%%%

We can now state our prescription for choosing boundary conditions
to reproduce any prescribed set of solitons.  Set
\be\label{Wnot}
W_0(\a) = -\int_0^\a \b_s(\a)
\ee
This function is universal, in the sense that it is independent of our choice of
boundary
conditions. Now given any smooth function $\V(\a)$ with $\V(0)=0$,
we write $\V=W_0+W$ and take our
boundary conditions to be $\b=W'(\a)$. It follows immediately that the extrema
of $\V$ are in one-to-one correspondence with solitons that obey these 
boundary conditions:
\be
0 = \V'=W_0' + W' = -\b_s + \b
\ee So the extrema of $\V$ are precisely the points where
$\b_s=\b$. Furthermore, the mass of each soliton is given by the
value of $\V$ at the corresponding extremum. To see this remember
that static solutions are extrema of the mass
\cite{Sudarsky:1992ty}.  Suppose we choose our boundary condition
to be $\b = \b_s(\a)$. For this special case, all the solitons are
allowed by the boundary conditions. Since we have a one parameter
family of static solutions, the mass must be constant, i.e., all
the solitons have the same mass. But this includes $\b=\a =0$
which is just AdS and has zero mass. So all the solitons have zero
mass. From (\ref{mass}), with boundary conditions $\b = \b_s(\a)$
we have \be 0 = M_0 + \a \b_s -W_0 \ee Therefore, for our general
boundary condition $\b=W'(\a)$, we have 
\be
M= 4\pi(M_0 + \a\b + W) =4\pi(W_0+W) =\oint \V d\Omega 
\ee
where we have used the fact that $\b = \b_s(\a)$ for
a soliton. Thus the mass of the soliton is indeed given by the
value of $\V$ at the corresponding extremum. Notice that the only 
restriction on $\V$ (that $\V(0)=0$) comes from the fact that we want the
total mass of pure AdS to be zero.

We have also studied the stability of these solitons. The most likely
mode to go unstable is a spherically symmetric scalar perturbation
like the one studied for hairy black holes in \cite{Maeda04}. We
have found numerically that this mode is indeed unstable if
$\V''<0$. We expect the solitons with $\V''> 0$ to be stable.
This leads to a new class of ``positive" energy conjectures
\footnote{This does not contradict our
earlier result (discussed in hep-th/0406134) that theories with solitons must 
have solutions with arbitrarily negative energy, since that 
assumed conventional boundary conditions that are scale invariant.
 The boundary conditions used here are generally
not invariant under rescaling $r$.}. For given boundary conditions, the
minimum energy solution is expected to be static, and hence one of
the solitons we have been discussing. If $\V$ has a global
minimum, then it seems likely that the energy of any
supergravity solution cannot be less than the minimum mass soliton.
In other words,

{\it Conjecture:  Given any smooth function $\V(\a)$  with $\V(0)=0$ 
and a global
minimum $\V_{min}$, consider  solutions to (\ref{4-action}) with
boundary condition $\b=W'$ where $W=\V-W_0$ and $W_0$ is given by
(\ref{Wnot}). Then the conserved mass (\ref{scalarst}) of
any nonsingular initial data set is bounded below by
$4\pi \V_{min}$.}

{\bf Field theory.}
We now turn to the dual field theory interpretation. String theory
on spacetimes which asymptotically approach $AdS_4\times S^7$ is
dual to the 2+1 conformal field theory (CFT)
describing the low energy excitations of a
stack of M two-branes. This theory is not well understood, but we
can learn something nontrivial using the gravitational solitons.
With $\b=0$ boundary conditions, the bulk scalar $\phi$ is dual to
a dimension one operator $\O$. One way of obtaining this CFT is by
starting with the field theory on a stack of D two-branes and
taking the infrared limit. In that description \cite{Aharony:1998rm},
\be \O = Tr T_{ij}\vp^i\vp^j \ee
where $T_{ij}$ is symmetric and traceless and
$\vp^i$ are the adjoint scalars.

Let $S_0$ denote the CFT lagrangian and  consider the deformation
\be\label{deform}
S= S_0 - k\int \O \ee
Using the standard AdS/CFT dictionary,
the vacuum expectation value of $\O$ in this deformed theory is
obtained by finding nonsingular static supergravity solutions with
$\b=-k$. But these are precisely our solitons. Given a soliton
with $\b=-k$, one has $\langle \O\rangle = \a$. Hence the function 
$\V(\alpha)=W_0-k\a$ can be interpreted as
the effective potential for $\langle \O\rangle$, where
$W_0$ is the function (\ref{Wnot}) computed earlier from the
soliton solutions.
From Fig. 1 we see that
there  are three qualitatively different regions. For small $k$,
there is a unique soliton and hence a unique nonzero value for 
$\langle \O\rangle$. For intermediate values of $k$ there are two
solitons indicating there are two vacua, and for large $k$ there
are no solitons indicating there is no vacuum at all.

Since our CFT lives on $S^2\times R$ (it is dual to a
bulk theory which approaches global AdS asymptotically) one might
have expected a mass term ${1\over 2}m^2 \vp^2$ coming from the
conformal coupling of the scalars to the curvature of the $S^2$.
The radius of the $S^2$ is equal to the AdS radius, so one
expects $m^2 = R/8 = 1/4$ in AdS units. Since $\O$ is quadratic in
$\vp$ the presence of a mass term would mean that for small 
$k<m^2/2$, the vacuum would be unchanged and $\langle \O\rangle=0$. 
But this is not what we find. Figure 1 shows that $\b_s$ is linear
in $\a$ for small $\a$, so that $W_0$ is quadratic in $\a$, hence the vacuum
expectation value $\langle \O\rangle$ is shifted even for small $k$.
This is illustrated in Figure 2 for $k=1/2$. For slightly larger $k$ a new 
maximum appears at larger $\a$ and the theory becomes nonperturbatively 
unstable.

%%%%%%%%%%%%%%%%%%%%%%%%%%%%%%%
\begin{figure}[htb]
\begin{picture}(0,0)
\put(30,140){$\V$} 
\put(210,50){$\alpha$}
\end{picture}
\psfig{file=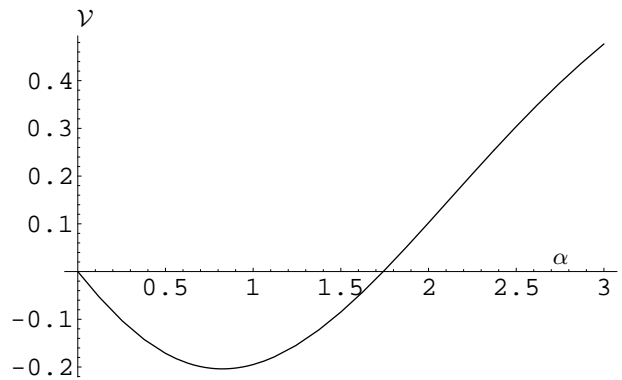,width=3.2in} 
\caption{The effective potential $\V (\a)=W_0 -\frac{1}{2}\a$.} 
\label{2}
\end{figure}
%%%%%%%%%%%%%%%%%%%%%%%%%%%%%%%%%%%

Now suppose we replace $-k\int \O$ in (\ref{deform}) with $\int
W(\O)$ where $W$ is an arbitrary function of $\O$. Remarkably, the
expectation values $\langle \O \rangle$ in different vacua are again given by the 
extrema of $\V=W_0+W$, where $W_0$ is the same function 
as above, and $W$ is unchanged.
This is because the addition of $\int W(\O)$ to the CFT action
corresponds in the bulk to using the modified boundary conditions
$\b=W'$ \cite{Witten02}. But we have already seen that the extrema
of $\V$ correspond to solitons with  precisely these boundary
conditions. The fact that the function $W$ does not receive any
corrections in the effective potential is surprising and
reminiscent of a nonrenormalization theorem, but we are dealing
with configurations that are far from supersymmetric. Perhaps it
is related to taking the large $N$ limit or to properties of
the operators that are dual to scalars with masses in the range (\ref{bound}).
An example of the effective potential in the presence of a 
multitrace deformation $W$ that yields a 
nontrivial false vacuum is given in Figure 3.

%%%%%%%%%%%%%%%%%%%%%%%%%%%%%%%
\begin{figure}[htb]
\begin{picture}(0,0)
\put(75,135){$\V$} 
\put(210,30){$\alpha$}
\end{picture}
\psfig{file=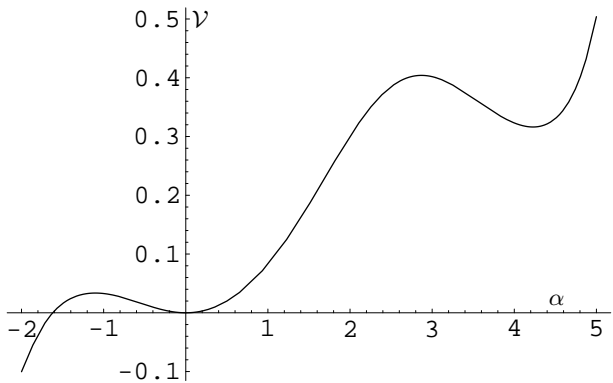,width=3.2in} 
\caption{The effective potential $\V = W_0 -\frac{1}{4}\a^2 +
\frac{1}{40}\a^3$.} 
\label{3}
\end{figure}
%%%%%%%%%%%%%%%%%%%%%%%%%%%%%%%%%%%

{\bf Discussion.}
In summary, we have seen that one can ``pre-order" solitons in supergravity,
in the following sense: 
Given essentially any function $\V(\a)$ there are boundary conditions such that gravitational
solitons exist precisely for each extremum of $\V(\a)$ and have masses given by the value of 
$\V$ at the corresponding extremum.  Furthermore, in supergravity theories with a field
theory dual, the function $\V$ can be interpreted as the effective potential for the dual
operator $\O$.
It would be interesting to perform an independent field theory calculation of the
effective potential, for instance in the case of a simple single trace deformation, since
this would provide a new test of AdS/CFT. 
One can also extend our results from solitons to black holes with scalar hair. 
One can pre-order black holes either in terms of their size or temperature. In the latter case,
a similar bulk analysis yields again a function $\tilde \V (\a)$ that can be interpreted as the finite 
temperature  effective potential  for $\O$ in the dual field theory.
This will be discussed in more detail in \cite{Hertog05}.

Although we have focused on a scalar field with $m^2=-2$ in four dimensional  $\N=8$ supergravity, the gravity side of the story can be generalized to other dimensions and all scalars with
masses in the range (\ref{bound}). For
 asymptotically $AdS_{d+1}$ spacetimes, a scalar field with mass $m$ asymptotically falls off like
 \be \label{asympt-sc}
\phi = {\alpha\over r^{\Delta_-}} + {\beta\over r^{\Delta_+}}
\ee
where
\be\label{roots}
\Delta_\pm = {d \pm \sqrt{d^2 + 4 m^2}\over 2}
\ee
 If the mass is
 in the range $m^2_{BF} \le m^2 < m^2_{BF} +1$, then a finite, conserved
 total energy
 can be defined for any boundary condition $\b(\a)$. The variation of the scalar
surface term is still (\ref{deltaQ}), so inserting this asymptotic
behavior of $\phi$ yields
\be \label{genmass}
 M=\mathrm{Vol}(S^{d-1})\left[ {d-1 \over 2} M_0 + \Delta_- \a\b +
(\Delta_+ - \Delta_-) W\right]
\ee
One can again construct the
soliton curve $\beta_s(\a)$ and find boundary conditions that admit
any desired soliton solutions.  If a dual field theory exists, then one
can again compute effective potentials for the operators dual to
the bulk scalar field.
\newline
{\bf Acknowledgments} It is a pleasure to thank O. Aharony,  J.
Maldacena, D. Marolf, J. Polchinski and N. Seiberg for discussions.
Part of this work was
done while G.H. was visiting the IAS in Princeton and he thanks
them for their hospitality. T.H. thanks the Solvay Institute at the
Universit\'e Libre de Bruxelles 
for its kind hospitality during the completion of this work.
This work was supported in part by NSF grant PHY-0244764.


\begin{thebibliography}{99}

\bibitem{Maldacena98}
J. M. Maldacena,
% ``The large N limit of superconformal field theories and supergravity,''
 Adv.\ Theor.\ Math.\ Phys.\  {\bf 2}
(1998) 231, hep-th/9711200.

\bibitem{Breitenlohner82}
P.~Breitenlohner and D.~Z.~Freedman,
% ``Stability In Gauged Extended Supergravity,''
 Annals Phys.\  {\bf 144} (1982) 249;
%``Positive Energy In Anti-De Sitter Backgrounds And Gauged Extended Supergravity,''
 Phys.\ Lett.\ B {\bf 115} (1982) 197.

\bibitem{Hertog04}
T. Hertog, K. Maeda,
%``Black Holes with Scalar Hair and Asymptotics in $N=8$ Supergravity,''
JHEP {\bf 0407} (2004) 051, hep-th/0404261.

\bibitem{Henneaux04}
M. Henneaux, C. Martinez, R. Troncoso, J. Zanelli,
%``Asymptotically Anti-de Sitter Spacetimes and Scalar Fields with a Logarithmic Branch,'' 
Phys. Rev. D {\bf 70} (2004) 044034, hep-th/0404236.

\bibitem{Duff99}
M. J. Duff, J. T. Liu,
%``Anti-de Sitter Black Holes in Gauged N=8 Supergravity,''
Nucl. Phys. {\bf B554} (1999) 237, hep-th/9901149.

%\cite{Klebanov:1999tb}
\bibitem{Klebanov:1999tb}
I.~R.~Klebanov and E.~Witten,
%``AdS/CFT correspondence and symmetry breaking,''
Nucl.\ Phys.\ B {\bf 556} (1999) 89,
hep-th/9905104.
%%CITATION = HEP-TH 9905104;%%

%\cite{Sudarsky:1992ty}
\bibitem{Sudarsky:1992ty}
D.~Sudarsky and R.~M.~Wald,
%``Extrema of mass, stationarity, and staticity, and solutions to the Einstein
%Yang-Mills equations,''
Phys.\ Rev.\ D {\bf 46} (1992) 1453.
%%CITATION = PHRVA,D46,1453;%%

%\cite{Maeda04}
\bibitem{Maeda04}
T.~Hertog and K.~Maeda,
%``Stability and thermodynamics of AdS black holes with scalar hair,''
hep-th/0409314.
%%CITATION = HEP-TH 0409314;%%

\bibitem{Aharony:1998rm}
O.~Aharony, Y.~Oz and Z.~Yin,
%``M-theory on AdS(p) x S(11-p) and superconformal field theories,''
Phys.\ Lett.\ B {\bf 430} (1998) 87,
hep-th/9803051;
%%CITATION = HEP-TH 9803051;%%
S.~Minwalla,
%``Particles on AdS(4/7) and primary operators on M(2/5) brane  worldvolumes,''
JHEP {\bf 9810} (1998) 002,
hep-th/9803053.
%%CITATION = HEP-TH 9803053;%%

\bibitem{Witten02}
E. Witten,
%``Multi-Trace Operators, Boundary Conditions, and AdS/CFTCorrespondence,'' 
hep-th/0112258.

\bibitem{Hertog05}
T. Hertog and G. T. Horowitz, to appear.

\end{thebibliography}
\end{document}